\documentstyle[sprocl,epsfig]{article}

\def\ra{\rightarrow}

\def\be{\begin{equation}}
\def\ee{\end{equation}}
\def\tmt{\times 10^{-2}}
\def\tmth{\times 10^{-3}}
\def\tmf{\times 10^{-4}}
\def\tmfv{\times 10^{-5}}
\newcommand{\beq}{\begin{equation}}
\newcommand{\eeq}{\end{equation}}
\newcommand{\bea}{\begin{eqnarray}}
\newcommand{\eea}{\end{eqnarray}}
\newcommand{\barr}{\begin{array}}
\newcommand{\earr}{\end{array}}
\newcommand{\bc}{\begin{center}}
\newcommand{\ec}{\end{center}}
\newcommand{\btab}{\begin{tabular}}
\newcommand{\etab}{\end{tabular}}
\newcommand{\gv}{\mbox{GeV}}
\newcommand{\tv}{\mbox{TeV}}

\newcommand{\nn}{\nonumber}

\newcommand{\dro}{\Delta\rho}
\newcommand{\drqcd}{\delta\!\rho\,_{QCD}}
\newcommand{\roro}{\rho^{(2)}}

\newcommand{\al}{\alpha}

\newcommand{\G}{\Gamma}
\newcommand{\Gmu}{G_{\mu}}

\newcommand{\ganu}{\gamma_{\nu}}

\newcommand{\gafi}{\gamma_5}

\newcommand{\noi}{\noindent}

\newcommand{\sm}{standard model }

\newcommand{\dal}{\Delta\alpha}

\newcommand{\mz}{M_Z^2}
\newcommand{\mw}{M_W^2}

\newcommand{\Dr}{\Delta r}

\newcommand{\alr}{A_{LR}}
\newcommand{\afb}{A_{FB}}

\newcommand{\ass}{asymmetries }

\newcommand{\pr}{{\it Phys.\ Rev.\ }}
 \newcommand{\prd}{{\it Phys.\ Rev.\ }{\bf D }}
\newcommand{\zp}{{\it Z.\ Phys.\ }{\bf C }}
\newcommand{\plb}{{\it Phys.\ Lett.\ }{\bf B }}
 \newcommand{\prl}{{\it Phys.\ Rev.\ Lett.\ }}
\newcommand{\np}{{\it Nucl.\ Phys.\ }{\bf B }}

\newcommand{\ms}{\overline{MS}}

\begin{document}

\title{ HIGGS MASS PREDICTION }

\author{ W. HOLLIK }

\address{Institut f\"ur Theoretische Physik, Universit\"at Karlsruhe, \\
D-76128 Karlsruhe, Germany}


\maketitle\abstracts{
In this talk the Higgs boson effects in electroweak precision observables
are reviewed and the possibility of indirect information on 
the Higgs mass from electroweak
radiative corrections and precision data is discussed.}

\section{Introduction}
By the present high  precision experiments
stringent tests on the standard model of electroweak and strong
interactions are imposed.
Impressive achievements have been made in the determination of the
$Z$ boson parameters \cite{blondel}, the $W$ mass
\cite{wmass}, and the confirmation
of the top quark at the Tevatron \cite{top,top1} with mass
$m_t = 175 \pm 6$ GeV, but   direct experimental evidence for the
Higgs boson is still lacking. 
 
Also a sizeable amount of theoretical work has
contributed over the last few years to a steadily rising
improvement of the standard model predictions
(for a review see ref.\ \cite{yb95}). The availability of both 
highly accurate measurements and theoretical predictions provides
tests of 
the quantum structure of the standard model thereby
probing the empirically yet unknown Higgs particle via its contribution
to the electroweak radiative corrections.

\section{Theoretical basis}
\subsection{Radiative corrections}

The possibility of performing precision tests is based
on the formulation of the \sm as a renormalizable quantum field
theory preserving its predictive power beyond tree level
calculations. With the experimental accuracy 
being sensitive to the loop
induced quantum effects, also the Higgs sector of the \sm
is probed. The higher order terms
induce the sensitivity of electroweak observables
to the top and Higgs mass $m_t, M_H$
and to the strong coupling constant $\al_s$, which are not present at the tree
level.

Before one can make predictions from the theory,
a set of independent parameters has to be taken from experiment.
For practical calculations the physical input quantities
$ \al, \; \Gmu,\; M_Z,\; m_f,\; M_H; \; \al_s $
are commonly used    
for fixing the free parameters of the standard model.
 Differences between various schemes are formally
of higher order than the one under consideration.
 The study of the
scheme dependence of the perturbative results, after improvement by
resumming the leading terms, allows us to estimate the missing
higher order contributions.
 
\smallskip
Two fermion induced
large loop effects in electroweak observables deserve a special
discussion:
\begin{itemize}
\item
The light fermionic content of the subtracted photon vacuum polarization
corresponds to a QED induced shift
in the electromagnetic fine structure constant. The recent update of the
evaluation of the light quark content
 \cite{eidelman,burkhardt}
 yield the result
\beq  (\dal)_{had} = 0.0280 \pm 0.0007\, . \eeq
Other determinations \cite{swartz}
agree within one standard deviation. Together with the leptonic
content, $\dal$ can
be resummed resulting in an effective fine structure
constant at the $Z$ mass scale:
\beq
   \al(\mz) \, =\, \frac{\al}{1-\dal}\,=\,
   \frac{1}{128.89\pm 0.09} \, .
\eeq
 \item
The electroweak mixing angle is related to the vector boson
masses  by
\bea
  \sin^2\theta  = 
   1-\frac{\mw}{\mz} + \frac{\mw}{\mz} \dro\, +  \cdots
  \, \equiv \,  s_W^2 + c_W^2 \dro\, +\cdots
\eea
where
the main contribution to the higher order quantity $\Delta\rho$
 is from the  $(t,b)$ doublet \cite{rho},
 in 1-loop and
 neglecting $m_b$ given by:
\beq
 \dro^{(1)} = 3 x_t, \;\;\;\; x_t =
 \frac{\Gmu m_t^2}{8\pi^2\sqrt{2}}
\eeq
Higher order
irreducible contributions have become available, modifying $\dro$
according to
 \beq
 \dro= 3 x_t \cdot [ 1+ x_t \,  \roro+ \drqcd ]
\eeq
 The electroweak 2-loop
 part \cite{bij,barbieri} is described by the
function $\roro(M_H/m_t)$ derived in \cite{barbieri} for general
Higgs masses.
$\drqcd$ is the QCD correction
to the leading $\Gmu m_t^2$ term
 \cite{djouadi,tarasov}
\beq
    \drqcd = - 2.86 a_s  - 14.6 a_s^2, \;\;\;\;
 a_s = \frac{\al_s(m_t)}{\pi} \, .
\eeq
The Higgs contribution to $\rho$ is only logarithmic for large Higgs masses.

\end{itemize}
\subsection{The vector boson masses}
The correlation between
the masses $M_W,M_Z$ of the vector bosons          in terms
of the Fermi constant $\Gmu$, in 1-loop order is given by
 \cite{sirmar}:
\beq
\frac{\Gmu}{\sqrt{2}}   =
            \frac{\pi\al}{2s_W^2 M_W^2} [
        1+ \Dr(\al,M_W,M_Z,M_H,m_t) ] \, .
\eeq
The decomposition
\beq
 \Dr = \Delta\al -\frac{c_W^2}{s_W^2}\,\dro^{(1)}
         + (\Dr)_{remainder} \, .
\eeq
separates the
leading fermionic contributions
                $\dal$ and $\dro$.
All other terms are collected in
the $(\Dr)_{remainder}$,
the typical size of which is of the order $\sim 0.01$.
 
\bigskip
The presence of large terms in $\Dr$ requires the consideration
of higher than 1-loop effects.
The modification of Eq.\ (7) according to
\bea
         1+\Dr & \, \ra\,& \frac{1}{(1-\Delta\al)\cdot
(1+\frac{c_W^2}{s_W^2}\dro) \, -\,(\Dr)_{remainder}}
 \nn \\
 & \equiv & \frac{1}{1-\Dr}
\eea
accommodates the following higher order terms
($\Dr$ in the denominator is an effective correction including
higher orders):
\begin{itemize}
\item
The leading log resummation \cite{marciano} of $\dal$:
$  1+\dal\, \ra \, (1-\dal)^{-1}$
\item
The resummation of the leading $m_t^2$ contribution \cite{chj}
in terms of $\dro$ in Eq.\ (5).
Beyond the $\Gmu m_t^2\al_s$ approximation through the $\rho$-parameter,
the complete
 $O(\al\al_s)$ corrections to the self energies
 are available from  perturbative  calculations
\cite{qcd} and by means of dispersion relations \cite{dispersion1}.
\item
With the quantity $(\Dr)_{remainder}$ in the denominator
non-leading higher order terms
containing mass singularities of the type $\al^2\log(M_Z/m_f)$
from light fermions
are also incorporated \cite{nonleading}.
\end{itemize}

\subsection{$Z$ boson observables}
With $M_Z$ as a precise input parameter, 
the predictions for the partial widths
as well as for the asymmetries
can conveniently be calculated in terms of effective neutral
current coupling constants for the various fermions.
The effective couplings follow
from the set of 1-loop diagrams
without virtual photons,
the non-QED  or weak  corrections.
These weak corrections
can be written
in terms of fermion-dependent overall normalizations
$\rho_f$ and effective mixing angles $s_f^2$
in the NC vertices:
\bea
 & &
 J_{\nu}^{NC}  =   \left( \sqrt{2}\Gmu\mz \right)^{1/2} \,
  (g_V^f \,\ganu -  g_A^f \,\ganu\gafi)  \\
 & &
   =  \left( \sqrt{2}\Gmu\mz \rho_f \right)^{1/2}
\left( (I_3^f-2Q_fs_f^2)\ganu-I_3^f\ganu\gafi \right)  . \nn
\eea
$\rho_f$ and $s_f^2$ contain  universal
parts     (i.e.\ independent of the fermion species) and
non-universal parts which explicitly depend on the type of the
external fermions.
The universal parts arise from the self-energies and contain the Higgs
mass dependence. The Higgs contributions to the non-universal
vertex corrections are suppressed by the small Yukawa couplings.

\smallskip
\paragraph{\it Asymmetries and mixing angles:}
 
The effective mixing angles are of particular interest since
they determine the on-resonance asymmetries via the combinations
   \beq
    A_f = \frac{2g_V^f g_A^f}{(g_V^f)^2+(g_A^f)^2}  \, .
\eeq
Measurements of the \ass hence are measurements of
the ratios
\beq
  g_V^f/g_A^f = 1 - 2 Q_f s_f^2
\eeq
or the effective mixing angles, respectively.

\smallskip
\paragraph{\it $Z$ width and partial widths:}
 
The total
$Z$ width $\Gamma_Z$ can be calculated
essentially as the sum over the fermionic partial decay widths
\beq
 \Gamma_Z = \sum_f \, \Gamma_f + \cdots , \;\;\;\;
 \Gamma_f = \Gamma  (Z\ra f\bar{f})
\eeq
The dots indicate other decay channels which, however,
are not significant.
 The fermionic partial
widths,
 when
expressed in terms of the effective coupling constants
read up to 2nd order in the fermion masses:
\bea
\Gamma_f
  & = & \G_0
 \, \left(
     (g_V^f)^2  +
     (g_A^f)^2 (1-\frac{6m_f^2}{\mz} )
                           \right)
 \cdot   (1+ Q_f^2\, \frac{3\al}{4\pi} ) 
          + \Delta\G^f_{QCD} \nn
\eea
with
$$
\G_0 \, =\,
  N_C^f\,\frac{\sqrt{2}\Gmu M_Z^3}{12\pi},
 \;\;\;\; N_C^f = 1
 \mbox{ (leptons)}, \;\; = 3 \mbox{ (quarks)}.
$$
and the QCD corrections  $ \Delta\G^f_{QCD} $
 for quark final states
 \cite{qcdq}.

\subsection{Accuracy of the standard model predictions}
 For a discussion of the theoretical reliability
of the \sm predictions one has to consider the various sources
contributing to their
uncertainties:

The experimental error of the hadronic contribution
to $\al(\mz)$, Eq.\ (2), leads to
$\delta M_W = 13$ MeV in the $W$ mass prediction, and
$\delta\sin^2\theta = 0.00023$ common to all of the mixing
angles, which matches with the experimental precision.

The uncertainties from the QCD contributions,
 besides the 3 MeV in the
hadronic $Z$ width, can essentially be traced back to
those in the top quark loops for the $\rho$-parameter.
They  can be combined into the following errors
\cite{kniehl95}:
 
$$
 \delta(\dro) \simeq 1.5\cdot 10^{-4},   \;
 \delta s^2_{\ell} \simeq 0.0001
$$
for $m_t = 174$ GeV.
 
The size of unknown higher order contributions can be estimated
by different treatments of non-leading terms
of higher order in the implementation of radiative corrections in
electroweak observables (`options')
and by investigations of the scheme dependence.
Explicit comparisons between the results of 5 different computer codes  
based on  on-shell and $\ms$ calculations
for the $Z$ resonance observables are documented in the ``Electroweak
Working Group Report'' \cite{ewgr} in ref.\ \cite{yb95}.
Table 1  shows the uncertainty in a selected set of
precision observables.
Quite recently (not included in table 1)
 the non-leading 2-loop corrections
$\sim \Gmu^2m_t^2 M_Z^2$ have been calculated \cite{padova}
for $\Delta r$ and $s_{\ell}^2$.
They reduce the uncertainty in $M_W$ and $s^2_{\ell}$ considerably,
by about a factor 0.2.

\begin{table}[htbp]\centering
\caption[]
{Largest half-differences among central values $(\Delta_c)$ and among
maximal and minimal predictions $(\Delta_g)$ for $m_t = 175\,\gv$,
$60\,\gv < M_H < 1\,\tv$, $\al_s(\mz) = 0.125$
(from ref.\ \cite{ewgr}) }
\vspace{0.5cm}
\begin{tabular}{c|c|c}
\hline \hline
Observable $O$ & $\Delta_c O$  & $\Delta_g O$ \\
\hline
            & & \\
$M_W\,$(GeV)          & $4.5\tmth$ & $1.6\tmt$\\
$\G_e\,$(MeV)          & $1.3\tmt$ & $3.1\tmt$\\
$\G_Z\,$(MeV)          & $0.2$     & $1.4$\\
$ s^2_e$             & $5.5\tmfv$ & $1.4\tmf$\\
$ s^2_b$             & $5.0\tmfv$ & $1.5\tmf$\\
$R_{had}$                 & $4.0\tmth$& $9.0\tmth$\\
$R_b$                 & $6.5\tmfv$ & $1.7\tmf$ \\
$R_c$                 & $2.0\tmfv$& $4.5\tmfv$ \\
$\sigma^{had}_0\,$(nb)    & $7.0\tmth$ & $8.5\tmth$\\
$\afb^l$             & $9.3\tmfv$ & $2.2\tmf$\\
$\afb^b$             & $3.0\tmf$ & $7.4\tmf$ \\
$\afb^c$             & $2.3\tmf$ & $5.7\tmf$ \\
$\alr$                & $4.2\tmf$ & $8.7\tmf$\\
\hline \hline
\end{tabular}
 
\end{table}

\section{Precision data and virtual Higgs bosons}
\begin{figure}[htb]
\vspace{-1cm}
\centerline{
\epsfig{figure=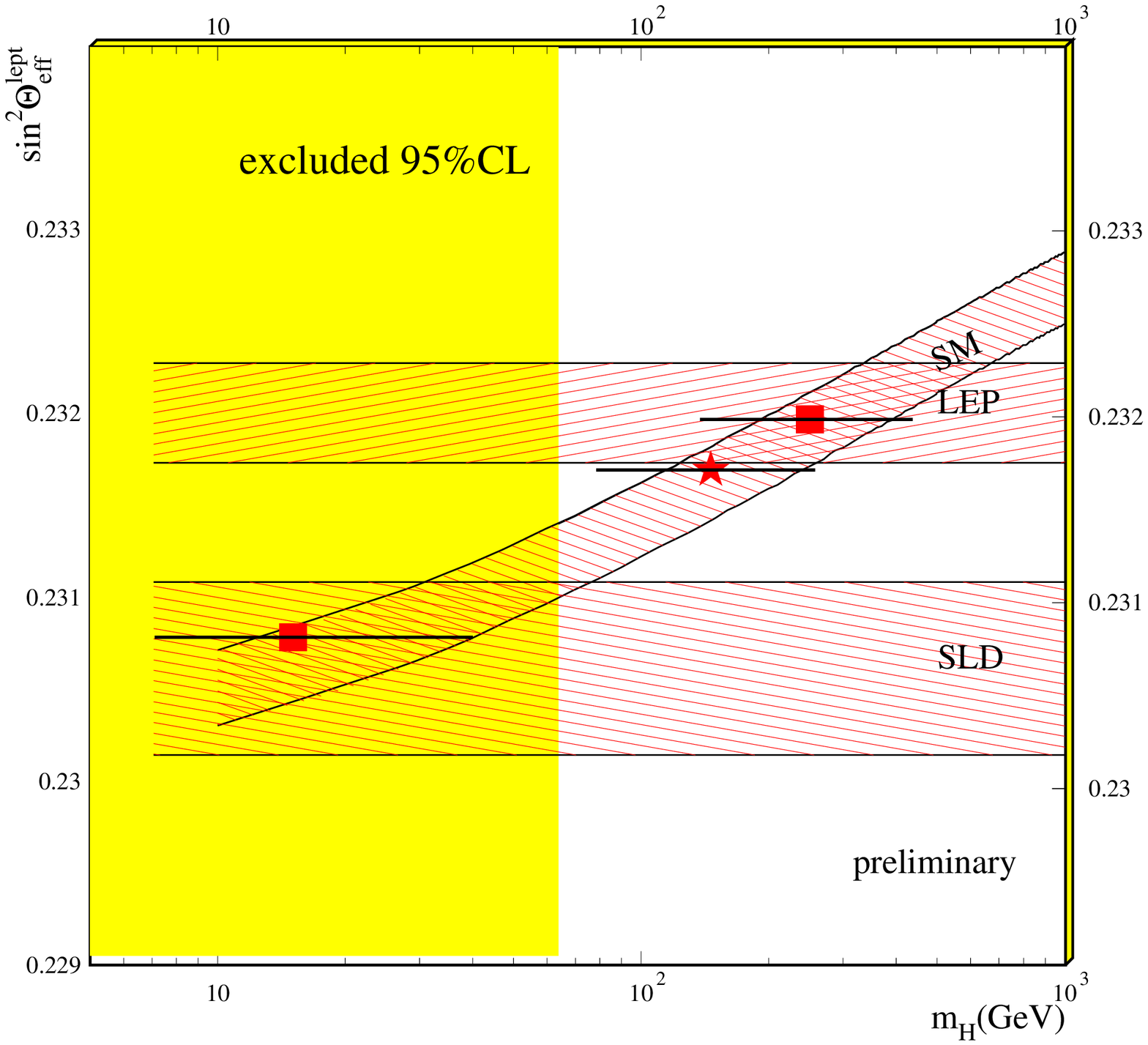,height=10cm}}
\vspace{-1.5cm}
\caption{Dependence of the leptonic mixing angle on the Higgs mass.
         The theoretical predictions correspond to
         $m_t=175\pm 6$ GeV. The SLD {\protect\cite{sld}}
         ($0.23061\pm 0.00047$) 
         and LEP {\protect\cite{blondel}} ($0.23200\pm 0.00027$)
         measurements are separately
         shown. The star is the result of a combined fit to LEP and SLD
         data, the squares are for separate fits 
         (from ref.\ {\protect\cite{deboer}}, updated version)} 
\label{semh}
\end{figure}

In table 2
the \sm predictions for $Z$ pole observables and the $W$ mass  are
put together for a light and a heavy Higgs particle with $m_t=175$ GeV.
 The last column is the variation of the prediction according to
$\Delta m_t = \pm 6$ GeV. The input value $\al_s = 0.123$
is the one from QCD observables at the $Z$ peak \cite{alfas}.
Not included are the uncertainties from
$\delta\al_s=0.006$, which amount to 3 MeV for the hadronic $Z$ width.
The experimental results on the $Z$ observables are from combined 
LEP and SLD data. $\rho_{\ell}$ and $s^2_{\ell}$ are the leptonic
neutral current couplings in eq.\ (10), derived from partial widths and
asymmetries  under the assumption of lepton universality.
The table illustrates the sensitivity of the various quantities 
to the Higgs mass.
The effective mixing angle turns out to be
the most sensitive  observable, where both the experimental error and the
uncertainty from $m_t$ are small compared to the variation with $M_H$.
Since a light Higgs boson goes along with a low value of $s^2_{\ell}$,
the strongest upper bound on $M_H$ is from $A_{LR}$ at the SLC \cite{sld},
 whereas 
LEP data alone allow to accommodate also a relatively heavy Higgs 
(see figure \ref{semh}).
Further constraints on $M_H$ are to be expected in the future from
more precise $M_W$ measurements at LEP 2.

\begin{table*}[t]
            \caption{Precision observables: experimental results 
             {\protect\cite{blondel}}
             and standard model         
             predictions. } \vspace{0.5cm}
            \bc
 \btab{| l | l | r | r | r | }
\hline
 observable & exp. (1996) &  $M_H=65$ GeV & $M_H=1$ TeV &
 $ \Delta m_t $ \\
\hline
\hline
$M_Z$ (GeV) & $91.1863\pm0.0020$ &  input & input &    \\
\hline
$\Gamma_Z$ (GeV) & $2.4946\pm 0.0027$ & 2.5015 & 2.4923 & $\pm 0.0015$ \\
\hline
$\sigma_0^{had}$ (nb) & $41.508\pm 0.056$ & 41.441 & 41.448 & $\pm 0.003$  \\
\hline
 $\G_{had}/\G_e$ & $20.778\pm 0.029 $ & 20.798 & 20.770 & $\pm 0.002$ \\
\hline
$\G_b/\G_{had}$  & $0.2178\pm 0.0011$ & 0.2156 & 0.2157 & $\pm 0.0002$ \\
\hline
$\G_c/\G_{had}$  & $0.1715\pm0.0056$ & 0.1724 & 0.1723 & $\pm 0.0001$ \\
\hline
$\rho_{\ell}$ & $1.0043\pm 0.0014$ & 1.0056 & 1.0036 & $\pm 0.0006$ \\
\hline
$s^2_{\ell}$  & $0.23165\pm 0.00024$ & 0.23115 & 0.23265 & $\pm 0.0002$ \\
\hline
$M_W$ (GeV) & $80.356 \pm 0.125$ & 80.414 & 80.216 & $\pm 0.038$  \\
\hline
\hline
\etab
\ec 
\clearpage
\end{table*}

Besides the direct measurement of the  $W$ mass,
the quantity $s_W^2$ resp.\  the ratio $M_W/M_Z$
is indirectly measured in deep-inelastic neutrino scattering,
in particular in the
NC/CC neutrino         cross section ratio for isoscalar targets.
The world average \cite{blondel} from CCFR, CDHS and CHARM,
including the new CCFR result
\cite{neutrino}
  $$ s_W^2 = 1-M_W^2/M_Z^2 = 0.2244 \pm 0.0044  $$
is fully consistent with the direct vector boson mass measurements.

\begin{figure}[htb]
\vspace{-1cm}
\centerline{
\epsfig{figure=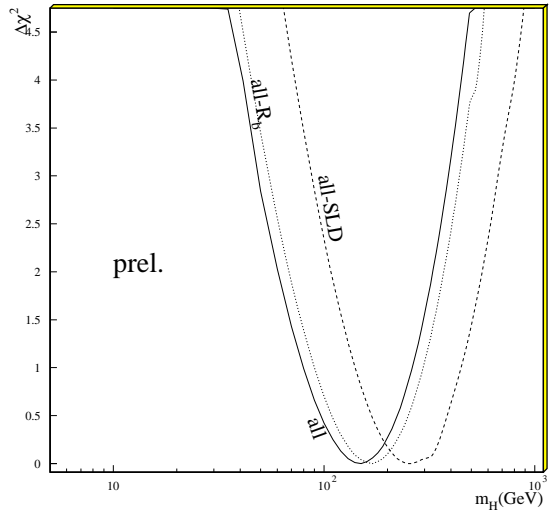,height=10cm}}
\vspace{-1.5cm}
\caption{Dependence of $\Delta\chi^2= \chi^2-\chi^2_{min}$
 on the Higgs mass (from ref.\ \protect \cite{deboer},
                     updated version)} 
\label{chi2mh}
\end{figure}

\paragraph{\it Standard model fits and Higgs mass range:}
Assuming the validity of the \sm a global fit to all electroweak results from
LEP, SLD, $M_W$,  $\nu N$ and $m_t$, 
allows to derive information on the allowed range for
the Higgs mass. Although the Higgs mass dependence of the electroweak
parameters is only logarithmic, the already quite accurate value for $m_t$
leads to some  sensitivity to $M_H$.  The Higgs mass dependence of
the $\chi^2$ of an overall fit is shown in figure \ref{chi2mh}
\cite{deboer}.
As one can see, the impact of $R_b$, which is on the way to the \sm value,
 is only marginal whereas $A_{LR}$
is decisive for a restrictive upper bound for $M_H$ (this is different from
the results based on the data from the last year \cite{higgs95}):

\smallskip \noi

including $\alr$:
\beq
  M_H = 146 ^{+112}_{-68} \gv, \;\;\;
  M_H < 364 \gv (95\% C.L.)  \nn
\eeq

\smallskip \noi
without $\alr$:
\beq
  M_H = 250 ^{+187}_{-112} \gv, \;\;\;
  M_H < 622 \gv (95\% C.L.)  \nn
\eeq

\smallskip \noi
Similar results have been obtained by Passarino \cite{passarino}.
The fit results by the LEP-EWWG \cite{blondel,gruenewald} are slightly higher
(see also \cite{jellis}):

\smallskip \noi
all data:
\beq
  M_H = 149 ^{+148}_{-82} \gv, \;\;\;
  M_H < 450 \gv (95\% C.L.)  \nn
\eeq

\smallskip \noi
Thee numbers do not yet include the theoretical uncertainties of the
standard model predictions. The LEP-EWWG \cite{blondel,gruenewald}
has performed a study of the influence of the various `options'
discussed in section 2.4 on the bounds for the Higgs mass with the result
that the 95\% C.L. upper bound is shifted by +100 GeV to higher values.
It has to be kept in mind, however, that this error estimate is based on the
uncertainties as given in table 1. Since the recent improvement in the 
theoretical prediction \cite{padova}
is going to reduce the theoretical uncertainty
especially in the effective mixing angle one may expect also a significant
smaller theoretical error on the Higgs mass bounds once the 2-loop terms
$\sim \Gmu^2 m_t^2 \mz$ are implemented in the codes used for the fits.
At the present stage the codes are without the new terms.

\section{Conclusions}
The quantum structure of the electroweak \sm allows in principle to probe
the mass of the as yet experimentally unknown Higgs boson through its
contribution to the radiative corrections for electroweak precision
observables. Although the dependence on $M_H$ is only logarithmic, the
experimental precision in the $Z$ boson parameters and the top quark mass
have meanwhile reached a level where a sensitvity to the Higgs mass
becomes visible, with pre\-ference to a light Higgs.
 The present upper bound on $M_H$ is dominated by the result on $\alr$.
The instability of the Higgs mass range obtained from global fits with or
without $\alr$ recommends to
consider the present mass bound with some caution. The only safe conclusion
is that we are well below the critical range where the standard Higgs 
becomes non-perturbative.
For the future, the reduction of the theoretical uncertainties and  more
precise experimental values for $M_W$ and $m_t$ will be the important
ingredients in improving the indirect Higgs search. 
 
\bigskip \noi
{\bf Acknowledgements:}
I want to thank W. de Boer, P. Gambino, M. Gr\"unewald, G. Passarino,
U. Schwickerath and G. Weiglein for helpful discussions and valuable
informations.


\section*{References}
\begin{thebibliography}{9}
\bibitem{blondel} A. Blondel, ICHEP96 (plenary talk), Warsaw, July 1996
\bibitem{wmass} M. Rijssenbeck, ICHEP96 (talk), Warsaw, July 1996
\bibitem{top} CDF Collaboration, F. Abe et al.,
              \prl {\bf 74} (1995) 2626;
             D0 Collaboration, S. Abachi et al., 
              \prl {\bf 74} (1995) 2632;
\bibitem{top1} P. Tipton, ICHEP96 (plenary talk); P. Grannis,
               ICHEP96 (talk), Warsaw, July 1996
\bibitem{yb95}  Reports of
             the Working Group on {\it Precision Calculations
             for the $Z$ Resonance}, CERN 95-03 (1995), Eds.\
             D. Bardin, W. Hollik, G. Passarino
\bibitem{eidelman} S. Eidelman, F. Jegerlehner,
                   \zp {\bf 67} (1995) 585
\bibitem{burkhardt} H. Burkhardt, B. Pietrzyk,
                    \plb {\bf 356} (1995) 398
\bibitem{swartz} M.L. Swartz, \pr {\bf D 53} (1996) 5268;
                 A.D. Martin, D. Zeppenfeld,
                 \plb {\bf 345} (1995) 558;
                 K. Adel, F.J. Yndurain, hep-ph/9509378;
                 D.H. Brown, W.A. Worstell, hep-ph/9607319
\bibitem{rho}
M. Veltman, \np  {\bf 123} (1977) 89;
M.S. Chanowitz, M.A. Furman, I. Hinchliffe, \plb {\bf 78} (1978) 285
\bibitem{bij} J.J. van der Bij, F. Hoogeveen, \np {\bf 283}
               (1987) 477
\bibitem{barbieri}
R. Barbieri, M. Beccaria, P. Ciafaloni, G. Curci, A. Vicere,
 \plb {\bf 288} (1992) 95; \np {\bf 409} (1993) 105;
J. Fleischer, F. Jegerlehner, O.V. Tarasov, \plb {\bf 319} (1993) 249
\bibitem{djouadi}
 A. Djouadi, C. Verzegnassi, \plb {\bf 195} (1987) 265
\bibitem{tarasov} L. Avdeev, J. Fleischer, S. M. Mikhailov, O. Tarasov,
                  \plb {\bf 336} (1994) 560;
                  E: \plb {\bf 349} (1995) 597;
                  K.G. Chetyrkin, J.H. K\"uhn, M. Steinhauser,
                  \plb {\bf 351} (1995) 331
\bibitem{sirmar} A. Sirlin, \pr {\bf 22} (1980) 971;
                 W.J. Marciano, A. Sirlin, \pr {\bf 22} (1980) 2695
\bibitem{marciano}
W.J. Marciano, \prd {\bf 20} (1979) 274
\bibitem{chj}
M. Consoli, W. Hollik, F. Jegerlehner, \plb {\bf 227} (1989) 167
\bibitem{qcd}
A. Djouadi, {\it Nuovo Cim.\ }{\bf A 100} (1988) 357;
D. Yu.\ Bardin, A.V. Chizhov, Dubna preprint E2-89-525 (1989);
B.A. Kniehl, \np {\bf 347} (1990) 86;
F. Halzen, B.A. Kniehl, \np {\bf 353} (1991) 567;
A. Djouadi, P. Gambino, \prd {\bf 49} (1994) 3499
\bibitem{dispersion1}
B.A. Kniehl, J.H. K\"uhn, R.G. Stuart, \plb {\bf 214} (1988) 621;
B.A. Kniehl, A. Sirlin, \np {\bf 371} (1992) 141;
                        \prd {\bf 47} (1993) 883;
 S. Fanchiotti, B.A. Kniehl, A. Sirlin, \prd {\bf 48} (1993) 307
\bibitem{nonleading}
A. Sirlin, \prd {\bf 29} (1984) 89
\bibitem{qcdq} For a review see:
  K.G. Chetyrkin, J.H. K\"uhn, A. Kwiatkowski, in [5], p.\ 175
\bibitem{kniehl95} B.A. Kniehl, in [5], p.\ 299
\bibitem{ewgr} D. Bardin et al., in [5], p.\ 7
\bibitem{padova} G. Degrassi, P. Gambino,
      A. Vicini, hep-ph/9603374; P. Gambino, private communication
\bibitem{alfas}
     S. Bethke, in: Proceedings of the  Tennessee International
                Symposium on
                Radiative Corrections, Gatlinburg 1994,
                Ed.\ B.F.L. Ward, World Scientific 1995
\bibitem{sld} E. Torrence (SLD Coll.), ICHEP96 (talk), Warsaw, July 1996
\bibitem{neutrino}
K. McFarlane (CCFR Coll.), ICHEP96 (talk), Warsaw, July 1996
\bibitem{deboer} W. de Boer, A. Dabelstein, W. Hollik, W. M\"osle,
                 U. Schwickerath, hep-ph/9607286,
                 updated version based on the data given at
                 ICHEP96, Warsaw, July 1996
\bibitem{higgs95} P. Chankowski, S. Pokorski, hep-ph/9509207;
                  J. Ellis, G.L. Fogli, E. Lisi, \zp {\bf 69} (1996) 627;
                  S. Dittmaier, D. Schildknecht, G. Weig\-lein,
                  hep-ph/9602436; 
                  G. Passarino, hep-ph/9604344 
\bibitem{passarino}
                 G. Passarino, talk at CRAD96, Cracow, August 1996
                 (updated from hep-ph/9604344)
\bibitem{gruenewald}
                 M. Gr\"unewald, ICHEP96 (talk), Warsaw, July 1996 
\bibitem{jellis} J. Ellis, G.L. Fogli, E. Lisi, hep-ph/9608329
 
\end{thebibliography}
\end{document}